\documentclass[3p,times]{elsarticle}

\usepackage{ecrc}

\usepackage{epstopdf}

\volume{00}

\firstpage{1}

\journalname{Nuclear Physics A}

\runauth{Chihiro Sasaki}

\jid{nupha}

\jnltitlelogo{Nuclear Physics A}

\usepackage{graphicx}
\usepackage{amsmath,amssymb}

\begin{document}

\begin{frontmatter}



\title{Selected Issues in Thermal Field Theory}

\author{Chihiro Sasaki}
\address{Frankfurt Institute for Advanced Studies,
D-60438 Frankfurt am Main,
Germany}
\address{Institute of Theoretical Physics, University of Wroclaw, 
PL-50204 Wroclaw, 
Poland}




\begin{abstract}
New developments on hot and dense QCD in effective field theories
are reviewed. Recent investigations in lattice gauge theories for
the low-lying Dirac eigenmodes suggest survival hadrons in restored
phase of chiral symmetry. We discuss expected properties of those bound 
states in a medium using chiral approaches. The role of higher-lying
hadrons near chiral symmetry restoration is also argued from
the conventional and the holographic point of view.
\end{abstract}


\end{frontmatter}



\section{Introduction}

The interplay between dynamical chiral symmetry breaking and color 
confinement in a hot/dense medium has not been sufficiently understood,
and remains one of the central subjects in QCD~\cite{review,qm:fuku}.
The chiral symmetry breaking and its restoration are well characterized
by the quark-antiquark (chiral) condensate, whereas no reliable order 
parameter for the confinement-deconfinement phase transition is known. 
The Polyakov-loop expectation value, which plays the role of the order 
parameter in pure Yang-Mills (YM) theory, is disturbed seriously by 
dynamical quarks. Hence, even though the expectation value exhibits an 
inflection point at a certain temperature, it is not manifest that 
the system undergoes a transition from hadrons to quarks and gluons.
A constructive way to identify the deconfined phase is to explore 
various fluctuations associated with conserved charges.
In particular, the kurtosis of net-quark number fluctuations measures 
clearly the onset of deconfinement~\cite{R42}.

\begin{figure}
\begin{center}
\includegraphics[width = 8cm]{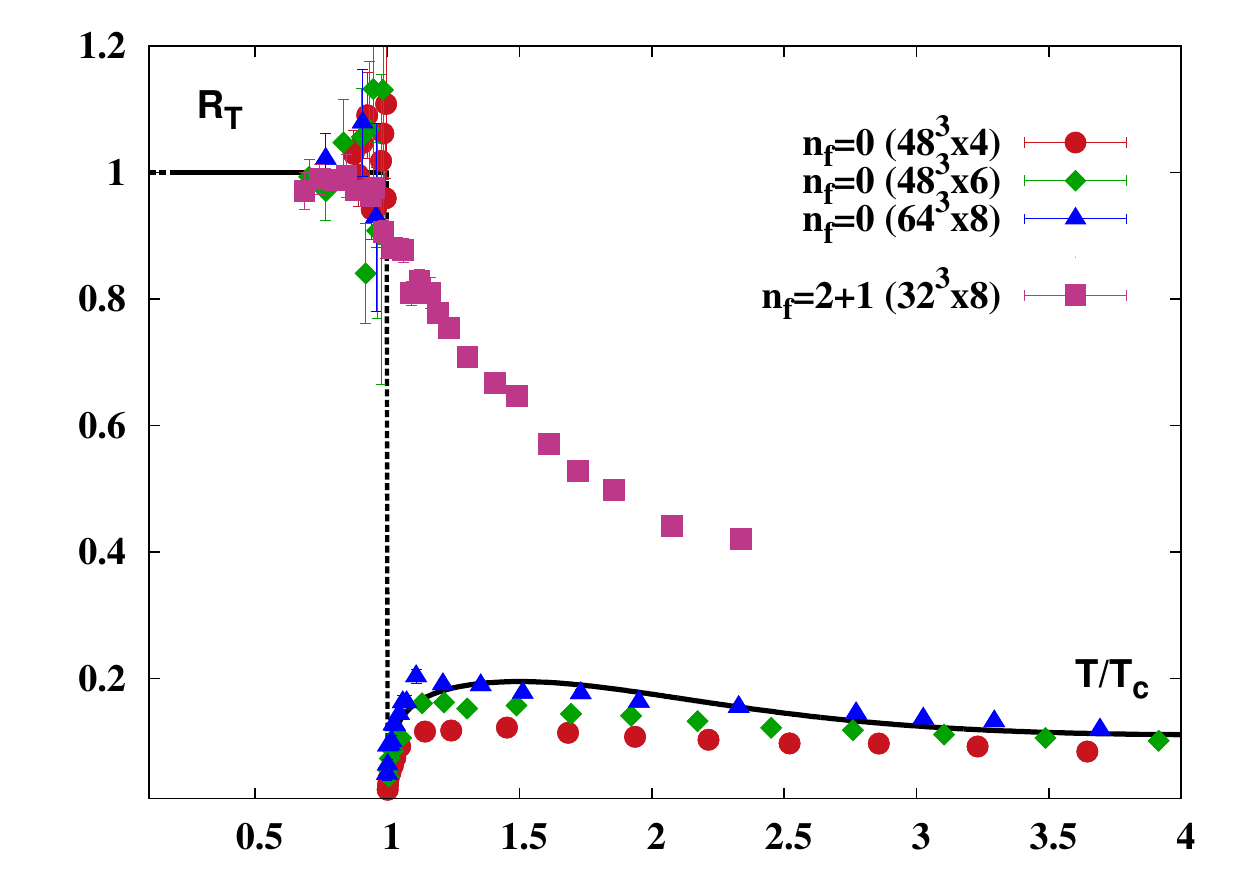}
\includegraphics[width = 8cm]{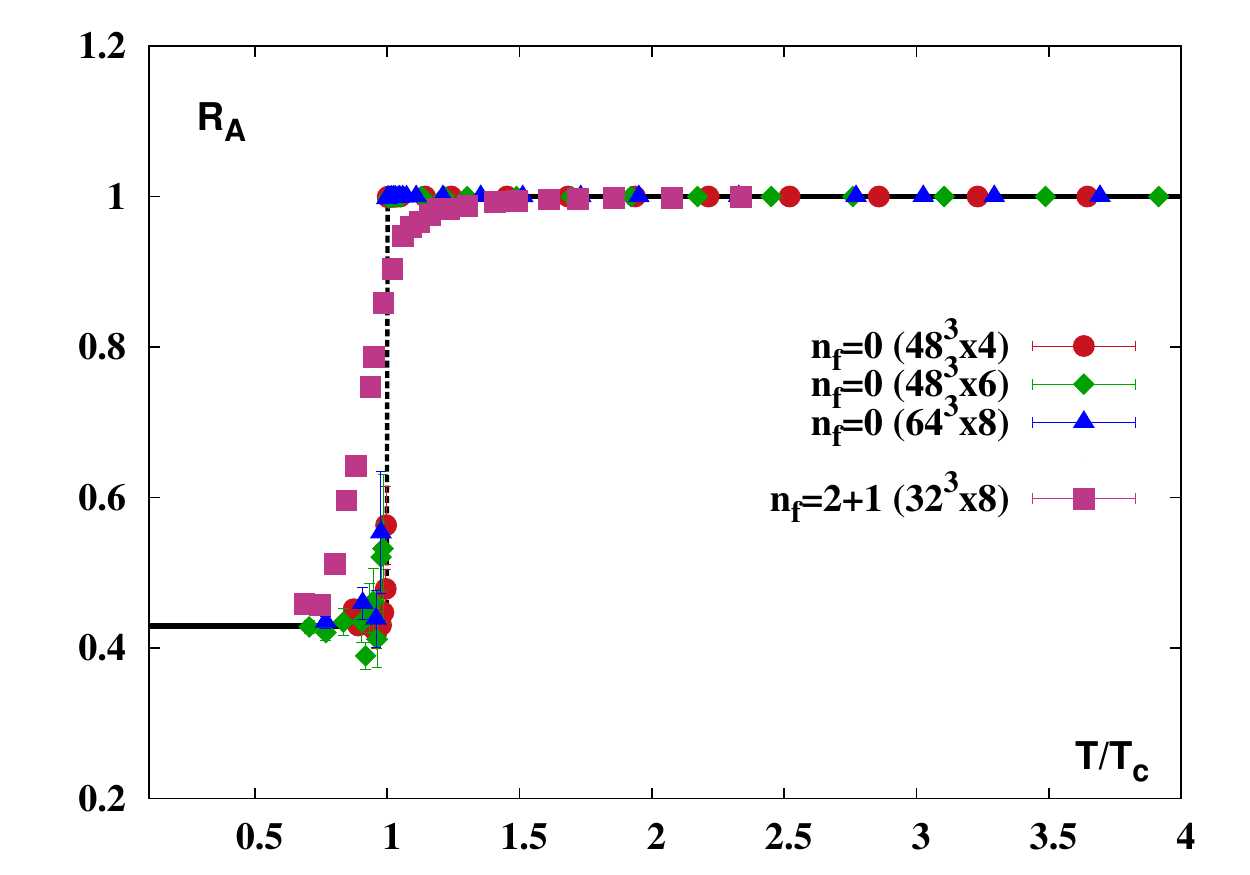}
\caption{
Lattice results of the ratios of the Polyakov loop susceptibilities,
$R_T = \chi_I/\chi_R$ and $R_A = \chi_A/\chi_R$ for pure YM and
$N_f=2+1$ QCD at vanishing chemical potential~\cite{Lo2}. 
The temperature is normalized by the critical temperature in pure YM 
theory, and by the pseudo-critical temperature for the chiral symmetry 
restoration in full QCD.
}
\label{RA}
\end{center}
\end{figure}

\begin{figure}
\begin{center}
\includegraphics[width = 12cm]{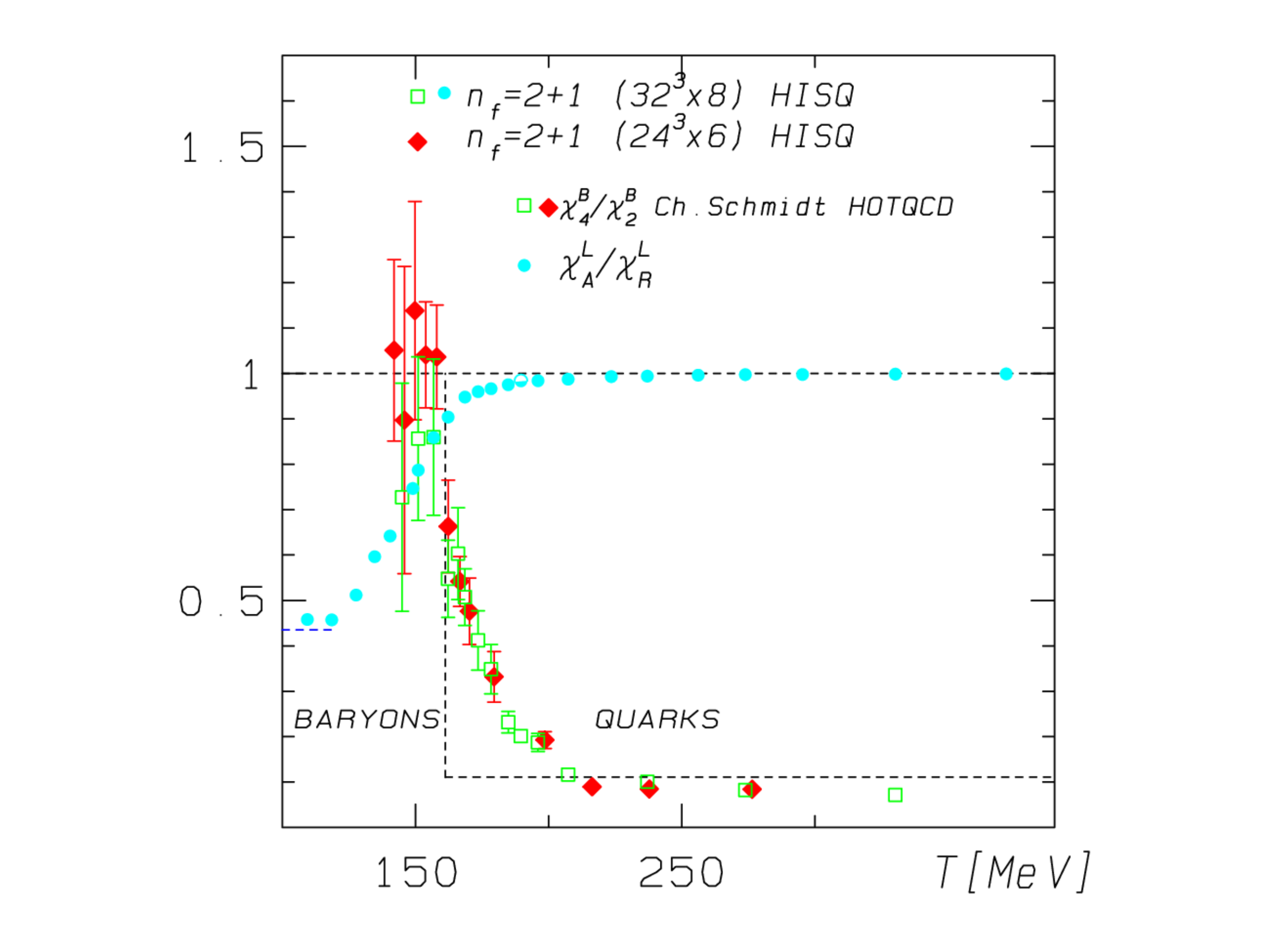}
\caption{
The ratio of the Polyakov loop susceptibilities $R_A=\chi_A/\chi_R$ 
and the kurtosis of net quark number fluctuations. Lattice data points
are taken from~\cite{Lo2,lat}.
}
\label{qf}
\end{center}
\end{figure}

Recently, other fluctuations more addressing the gluon sector have been 
calculated in lattice gauge theory with light quarks~\cite{Lo1,Lo2}, where 
two ratios, $R_T = \chi_I/\chi_R$ and $R_A = \chi_A/\chi_R$, are considered
in terms of the susceptibilities associated with the modulus, real and 
imaginary parts of the Polyakov loop.
Asymptotic values of those ratios are properly quantified within a 
$Z(3)$-symmetric model when there are no dynamical quarks (see 
Fig.~\ref{RA}). Once the light flavored quarks are introduced, the $R_T$ 
becomes much broadened, similarly to the Polyakov loop expectation value.
On the other hand, the $R_A$ retains the underlying center symmetry
fairly well even in full QCD with the physical pion mass. Also, ambiguities
of the renormalization prescription can be avoided to large extent in
the ratio. The $R_A$ thus serves as a better pseudo-order parameter
than the Polyakov loop by itself. 
In Fig.~\ref{qf}, the $R_A$ is compared with the kurtosis of the quark 
number fluctuations. The quark liberation takes place evidently together
with a qualitative changeover in $R_A$. Those abrupt changes in the 
Polyakov loop and quark number fluctuations appear in a narrow range 
of temperature lying on the pseudo-critical temperature of chiral symmetry
restoration. Therefore, at vanishing chemical potential, 
$T_{\rm deconf} \simeq T_{\rm chiral}$ is concluded.

{}From the field theoretical point of view, it remains incomplete to
capture the interplay of such non-perturbative dynamics in a form of 
an effective theory. In this contribution, we will briefly review recent 
progress in QCD thermodynamics and address the issues to be disentangled.

\section{Low-lying Dirac eigenmodes and confinement}

\begin{figure}
\begin{center}
\includegraphics[width = 10cm]{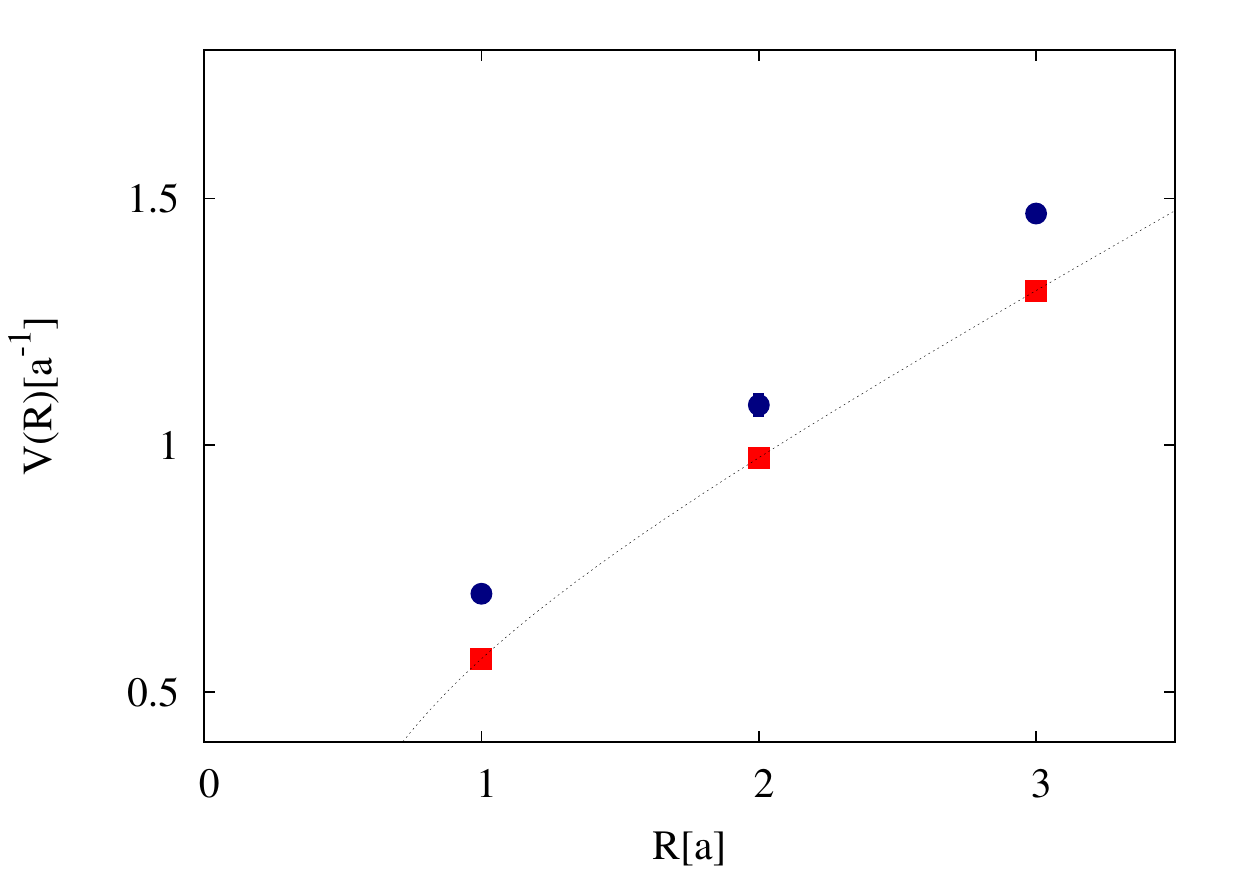}
\caption{
Static quark potential in SU(3) lattice gauge theory~\cite{Gongyo}.
The points with filled circles (squares) represent the results
with (without) removal of the low-lying Dirac modes.
}
\label{dirac:pol}
\end{center}
\end{figure}

Spontaneous chiral symmetry breaking is locked to a non-vanishing 
spectral density with the zero eigenvalues of the Dirac operator, 
known as the Banks-Casher relation~\cite{BC}. An intriguing question
is whether confinement would also be lost if the Dirac zero modes are 
artificially removed from a system. 
In~\cite{Gattringer,Bruckmann,Synatschke,Gongyo,Doi}, a relation of 
the Dirac eigenmodes to the Polyakov loop has been formulated on a lattice,
and their dynamical correlations have been investigated in SU(3) lattice
gauge theory. The expression in a gauge-invariant formalism is found 
as~\cite{Gongyo,Doi}
\begin{equation}
\langle L \rangle
= \frac{(2i)^{N_\tau-1}}{12V}\sum_n \lambda_n^{N_\tau-1}
\langle n | \hat{U}_4 | n \rangle\,.
\label{eq:poly}
\end{equation}
Those simulations revealed that there are no particular
modes which crucially affect confinement. In fact, the string tension
extracted from the potential between static quarks is unchanged even
when the low-lying Dirac modes are eliminated, as shown in 
Fig.~\ref{dirac:pol}. This apparently indicates that the disappearance
of the chiral symmetry breaking does not dictate deconfinement of quarks.
The derived analytic relation (\ref{eq:poly}) tells manifestly that the 
Dirac zero modes has no role in the Polyakov loop. However, the matrix 
elements of the link variable must be affected by the light quarks, so that 
the Wilson loop and the quark potential may be significantly modified.
The simulations carried out so far are limited to the quenched theories.
Therefore, a conclusive statement should be postponed until computations
with the dynamical quarks are made in future.

A conventional picture for color confinement is based on the dual
superconductor~\cite{mono1,mono2}. With a particular gauge fixing for
the QCD Lagrangian, so-called Maximum Abelian Gauge~\cite{MAG1,MAG2}, 
magnetic monopoles naturally emerge and get condensed. As a result, 
a linear confinement potential and dynamical chiral symmetry breaking 
are generated. In order to accommodate the relation between the Polyakov
loop and Dirac zero eigenmodes into this scenario, one needs to somehow
link the Dirac eigenmodes to (a part of) the monopoles.

\section{Hadrons near chiral symmetry restoration}

\begin{figure}
\begin{center}
\includegraphics[width = 12cm]{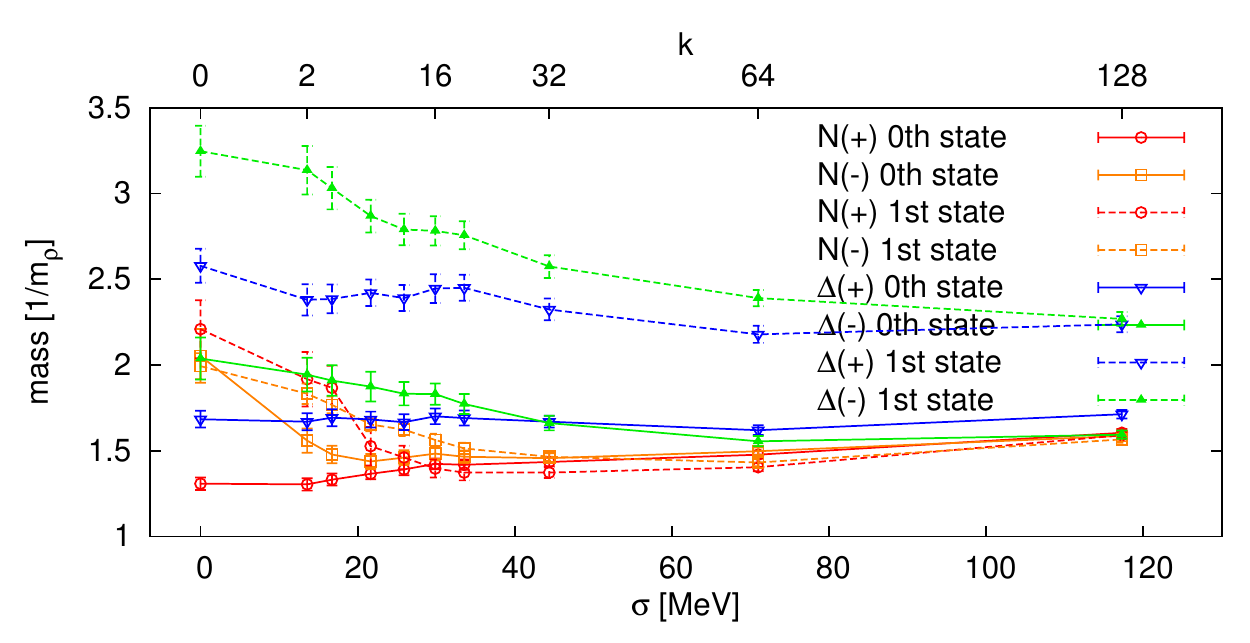}
\includegraphics[width = 12cm]{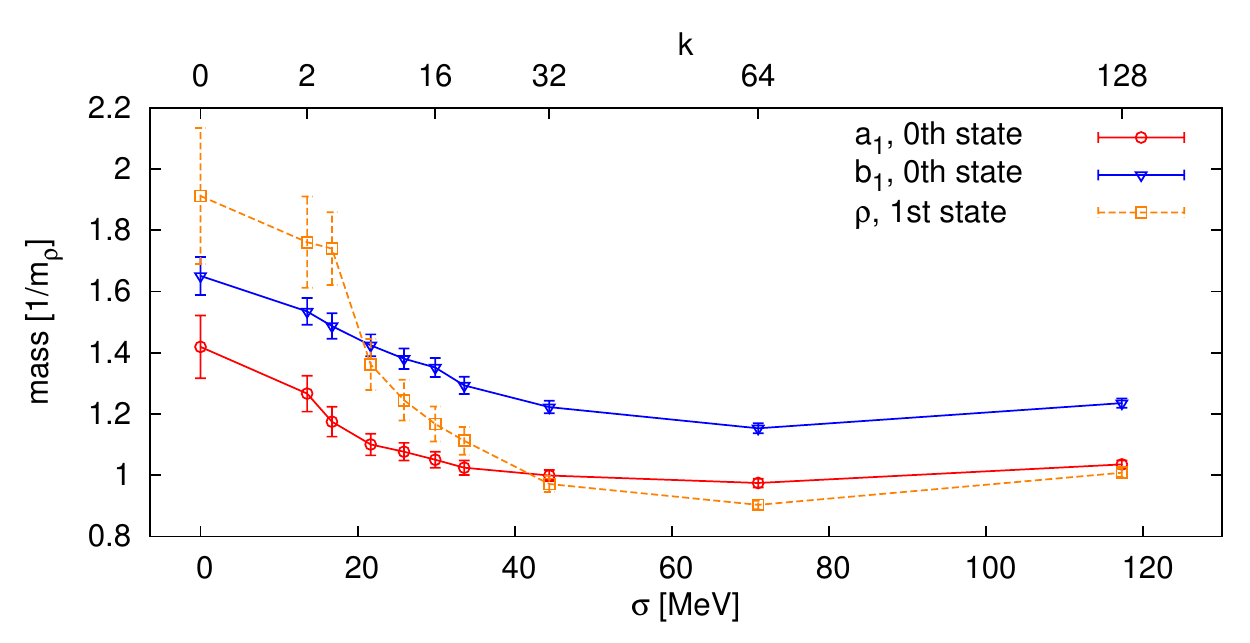}
\caption{
Masses of baryons (upper) and mesons (lower) in unit of the $\rho$
meson mass as a function of the truncation level~\cite{GLS}.
}
\label{dirac:mass}
\end{center}
\end{figure}

Another implication of the chiral symmetry restoration with confinement
is found in the hadron mass spectra with a systematic removal of
the low-lying Dirac modes on a lattice~\cite{GLS}. The masses of several
baryons and mesons with positive and negative parity are summarized 
in Fig.~\ref{dirac:mass}.
As increasing the truncation level of the Dirac modes, the masses of parity
partners approach and eventually become degenerate. What is remarkable is
that those hadrons remain quite massive, around 1 GeV for the lowest nucleon
and $m_\rho$ for the lowest vector meson. Furthermore, universal scaling
--- $2m$ for mesons and $3m$ for baryons --- is not observed.
Hence, it is considerably suggestive that those hadrons keep their particle
identities and survive in the chiral restored phase.

\begin{figure}
\begin{center}
\includegraphics[width = 10cm]{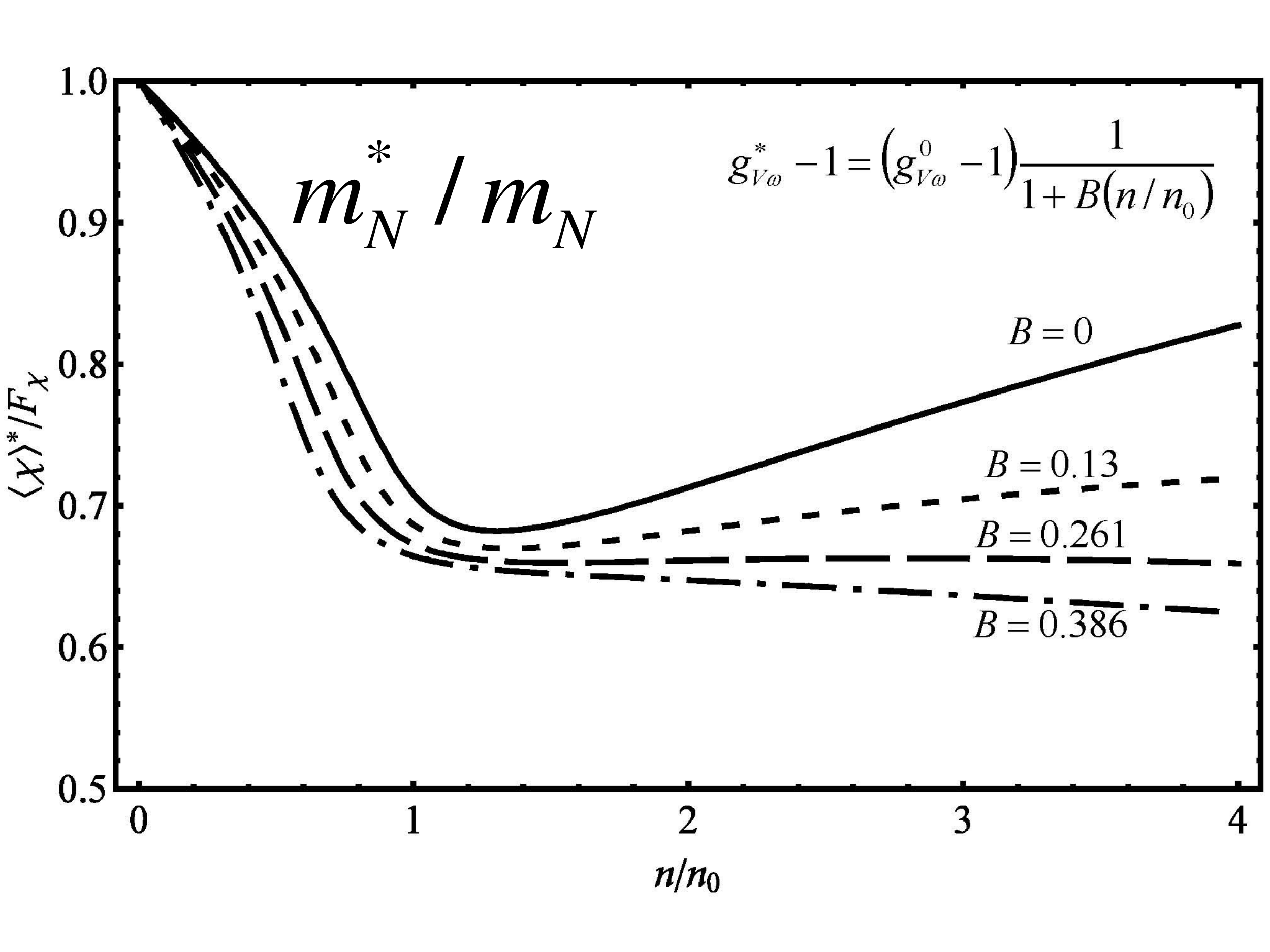}
\caption{
The ratio of the in-medium nucleon mass to its vacuum value
for the density-dependent omega-nucleon coupling $g_{V_\omega}$~\cite{PLRS}.
}
\label{dilaton}
\end{center}
\end{figure}

Given the lattice observation that $T_{\rm deconf} \simeq T_{\rm chiral}$
at vanishing chemical potential, the chiral restored phase with confinement
might appear at high density. A large hadron mass needs to be saturated by
certain condensates of chirally even operators. A good candidate is
gluon condensates. Not only in matter-free space but also in a medium, 
the QCD trace anomaly exists and this is accompanied by a non-vanishing
expectation value of a dilaton field, which is identified with a scalar 
glueball~\cite{Schechter}. The in-medium gluon dynamics in the context 
of scale symmetry breaking is accommodated in a chiral effective field 
theory. In~\cite{PLRS}, the $\rho$ and $\omega$ mesons are shown to interact
with a nucleon differently: the $\rho NN$ coupling runs, whereas the 
$\omega NN$ coupling walks in density. An immediate consequence is that 
the in-medium nucleon mass reaches a constant around the saturation 
density, and stays in higher density, as given in Fig.~\ref{dilaton}.
Note that the original Lagrangian does not have an explicit ``bare'' mass.
Nevertheless, due to the dynamics in dense matter, a chirally-invariant 
mass for the nucleon emerges. The above density dependences encoded in 
the renormalization group equations are governed by a non-trivial IR fixed 
point, dilaton-limit fixed point~\cite{DL1,DL2}. These features remind us of 
the modern technicolor models for the Higgs physics beyond the Standard 
Model~\cite{techni}.

Higher-dimension operators can also be condensed and contribute to the
hadronic quantities in dense matter. In particular, tetra-quark states
play a crucial role to construct reliable equations of state
for nuclear matter~\cite{Gallas} and near the chiral phase transition via
a mixing to a bilinear quark condensate~\cite{Heinz}. 
Also, a novel phase with chiral symmetry breaking on top of the vanishing 
chiral condensate is an interesting theoretical option at finite 
density~\cite{HST}, where the tetra-quark condensate saturates the pion 
decay constant and could yield more critical point(s) in the phase diagram.

\section{Role of higher-lying hadrons}

\begin{figure}
\begin{center}
\includegraphics[width = 14cm]{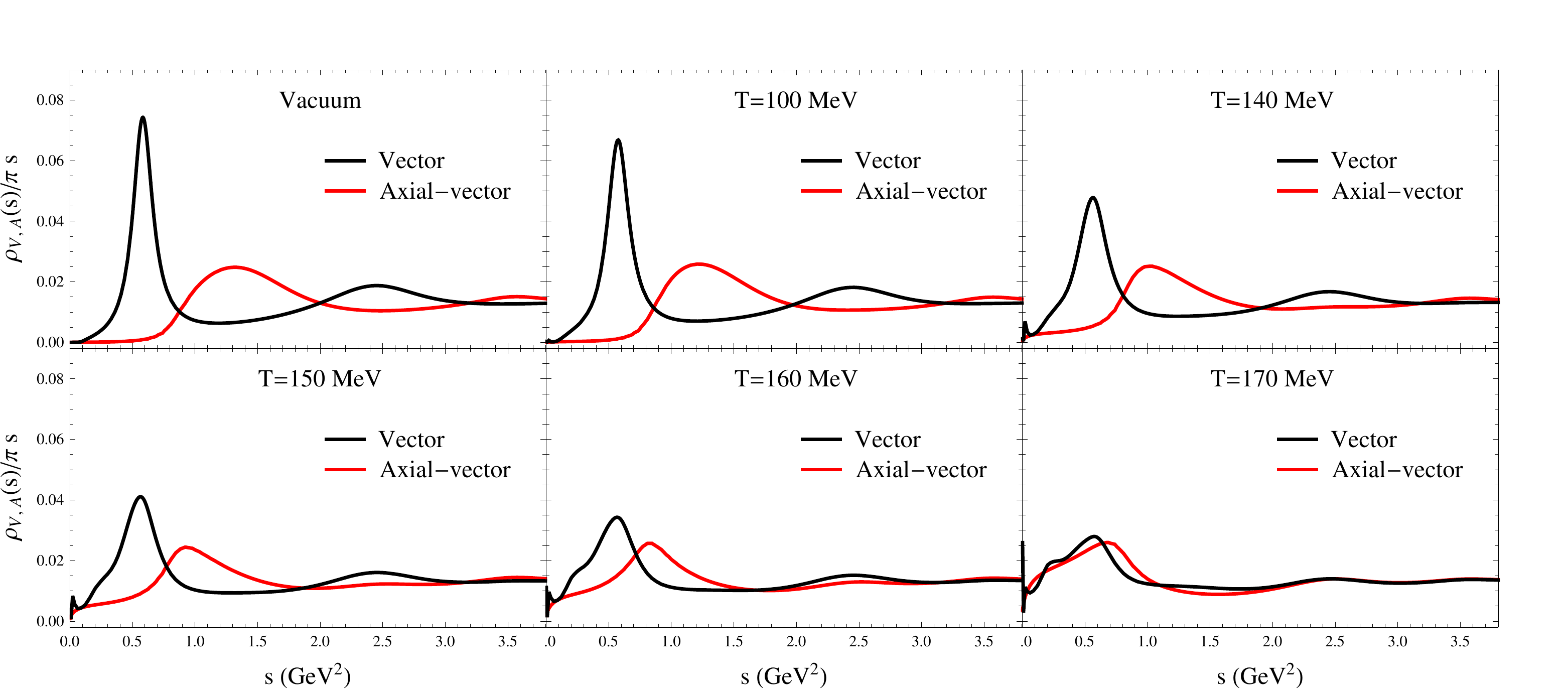}
\caption{
Vector and axial-vector spectral functions at 
various temperatures~\cite{HR}.
}
\label{va:t}
\end{center}
\end{figure}

The in-medium vector spectrum $\rho_V$ is more or less established both 
in theory and in dilepton measurements~\cite{RWvH}. Yet, it has not been 
clarified how the observed modifications are linked to the (partial) chiral
symmetry restoration. Instead of measuring the in-medium axial-vector 
spectrum $\rho_A$ in experiments, which is hopeless, a method to construct
$\rho_A$ using a phenomenologically accepted $\rho_V$ via QCD and Weinberg
sum rules has been proposed~\cite{HR}.
Fig.~\ref{va:t} summarizes the obtained thermal evolution of the spectra.
The $a_1$ meson mass smoothly approaches the $\rho$ mass, and the two
spectral functions become almost on top at a high temperature, indicating
chiral restoration. Not only the lowest vector states, $\rho$ and $a_1$,
but also the second lowest states, $\rho^\prime$ and $a_1^\prime$, are
shown to contribute to the $\rho_{V,A}$ rather significantly as approaching
the restoration temperature. It is intuitively understood since
more hadronic states must be populated toward the chiral phase
transition that cannot be achieved within any conventional perturbative 
treatment including just a few numbers of mesonic states.

\begin{figure}
\begin{center}
\includegraphics[width = 10cm]{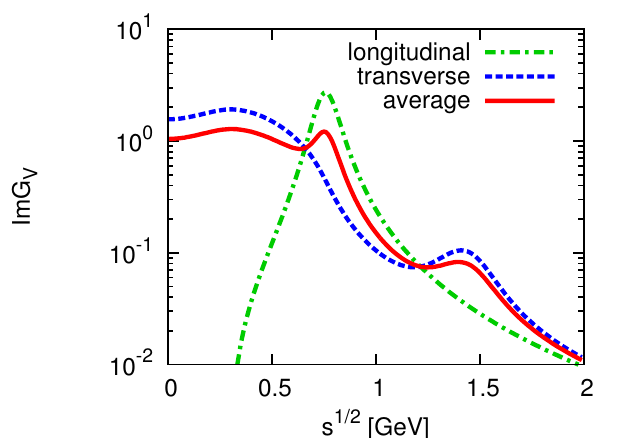}
\caption{
Vector spectral function for $C=1$ GeV~\cite{HS}.
}
\label{va:mu}
\end{center}
\end{figure}

At non-vanishing chemical potential, it is more involved since charge
conjugation invariance is lost, which leads to a mixing between transverse
$\rho$ and $a_1$ states at tree level: 
\begin{equation}
{\mathcal L}_{\rm mix}
= 2C\epsilon^{0\nu\lambda\sigma}
\text{tr}\left[
\partial_\nu V_\lambda\cdot A_\sigma
{}+ \partial_\nu A_\lambda\cdot V_\sigma
\right]\,.
\end{equation}
Their dispersion relations are
modified and the spectral functions do not follow a simple Breit-Wigner
distribution~\cite{HS} (see Fig.~\ref{va:mu}). 
Its relevance on observables crucially relies on a mixing strength $C$ 
which intrinsically depends on density. There are two available numbers:
$C = 1$ GeV at the saturation density $n_0$ from an AdS/QCD model~\cite{DH},
and $C = 0.1$ GeV at $n_0$ from the gauged Wess-Zumino-Witten action in 
four dimensions as well as a mean field approximation~\cite{HS}. The former
yields vector meson condensation slightly above $n_0$, which is odd. Therefore,
this is most likely an artifact of the large $N_c$ approximation employed
for the gauge/gravity duality conjecture.

One conceivable reason for this huge difference in C's is that all the
Kaluza-Klein (KK) modes, corresponding to all the vector mesons, contribute
to the dynamics in holographic QCD models. Heavier states can be integrated
out, whereas a naive truncation may provide a different result since 
truncated modes carrying the information about the underlying physics are
artificially omitted. In fact, the nuclear potential as a function of
a distance $r$ exhibits a $1/r^2$ dependence when all the KK modes are
considered~\cite{HSS}, while it follows a $1/r$ behavior when a truncation
is made.

As increasing temperature and density, more hadrons are activated
and eventually change the ground state. It is not straightforward to
deal with many (or all the) hadrons. In holographic QCD approach,
although infinite KK modes are naturally  accommodated, a systematic
technique to include $1/N_c$ corrections in a medium is not established yet.
In the standard effective theories in four dimensions, interactions with 
the higher-lying hadrons are not completely known. Integrating the heavier
modes out at finite temperature and density is not an easy task either. Those 
effective interactions near the phase transition may be to some extent
captured by use of more microscopic computations, e.g. lattice simulations,
Dyson-Schwinger equations and functional approach~\cite{qm:frg}.

\section{Conclusions}

Various fluctuations of conserved changes~\cite{qm:lat} as well as
the ratio of the Polyakov-loop susceptibilities $R_A$ in lattice QCD 
consistently indicate that deconfinement takes place in the chiral 
crossover region at vanishing chemical potential. Although the kurtosis of
net quark number fluctuations is reasonably quantified in a class of
chiral models with the Polyakov loop~\cite{pnjl}, modifications in
$R_A$ and $R_T$ by the dynamical quarks cannot be explained in the same
framework. Since those models do not posses the dynamical mechanism for 
quark confinement, the properties of gluon-oriented quantities are
supposed to be less captured. An effective theory that can better handle
the confinement nature of non-abelian gauge theories is indispensable
to reveal the Polyakov loop fluctuations in the presence of light quarks.
Also, it is vital to bridge the gap between the Dirac zero eigenmodes,
responsible for dynamical chiral symmetry breaking, and the magnetic monopoles.

The role of higher-lying hadrons has been found in low-energy constants and 
spectral functions near the QCD phase transition. On a practical level,
it is not yet established to fully accommodate them to effective theories.
Several attempts constrained by the relevant global symmetries lead to
certain non-trivial medium effects. More elaborated and systematic prescription
certainly requires a novel scheme. Functional approaches in terms of quarks
and gluons may provide some benefits in this context.

Heavy-light hadrons, such as charmed mesons, are also good probes for
the quark-gluon dynamics. In dilute nuclear matter, in-medium modifications
of the color-electric and color-magnetic gluons are extracted from the
D and B meson dynamics~\cite{YS}. In increasing density/temperature, those 
heavy-light mesons will change their chiral properties, as expected from
the chiral doubling scenario~\cite{cd1,cd2,cd3}. The mass gap between the
chiral partners is around 350 MeV, and this is much bigger than a mass
difference between the charged D mesons in dense matter, $\sim 50$ MeV.
Further theoretical investigations, along with the lattice input~\cite{Burger},
will supply more reliable understanding of heavy-flavor transport properties.

\begin{figure}
\begin{center}
\includegraphics[width = 10cm]{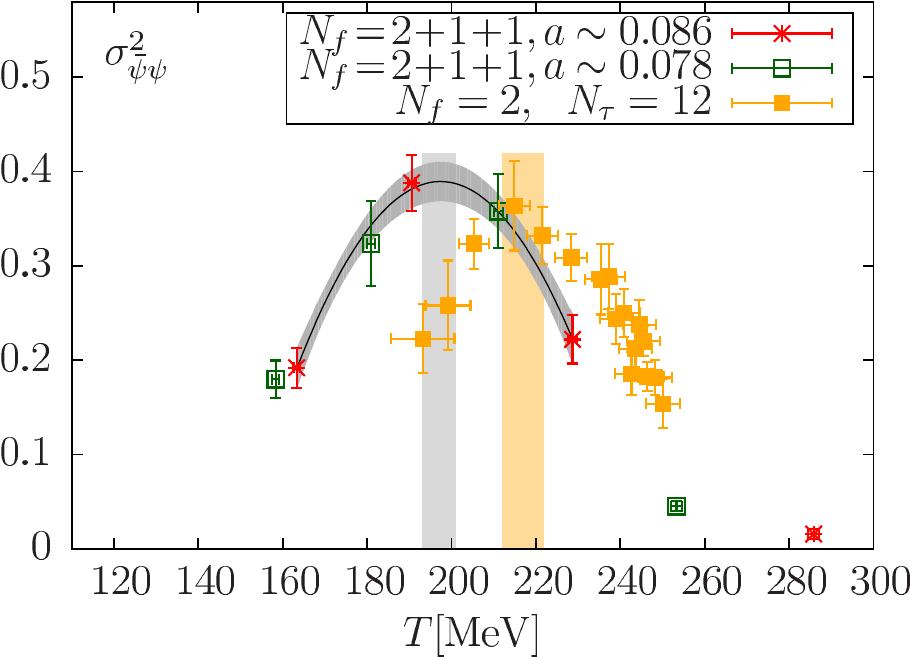}
\caption{
The chiral susceptibility calculated in lattice QCD 
for $N_f=2$ and $N_f=2+1+1$~\cite{Burger}.
}
\label{charm}
\end{center}
\end{figure}

\section*{Acknowledgments}

I am grateful for fruitful discussions and correspondence with
K.~Redlich, H.~Suganuma and S.~Sugimoto.
I acknowledge partial support by the Hessian
LOEWE initiative through the Helmholtz International
Center for FAIR (HIC for FAIR), and 
by the Polish Science Foundation (NCN) under
Maestro grant 2013/10/A/ST2/00106.








\end{document}